\documentclass[%
 aip,jcp,
preprint,%
]{revtex4-1}

\usepackage{overpic} 
\usepackage{braket}
\usepackage{subfigure}
\usepackage{graphicx}
\usepackage{xcolor}
\usepackage{dsfont}
\usepackage{pstricks}

\begin{document}


\preprint{AIP/123-QED}

\title{Modeling Charge Transport in $C_{60}$-based Self-assembled Monolayers for 
Applications in Field-Effect Transistors}




\author{S.\ Leitherer}
\affiliation{
Institute for Theoretical Physics and Interdisciplinary Center for Molecular Materials, \\
University Erlangen-N\"urnberg,\\
Staudtstr.\ 7/B2, D-91058 Erlangen, Germany
}
\author{C.\ M.\ J\"ager}
\affiliation{
Computer-Chemie-Centrum and Interdisciplinary Center for Molecular Materials,\\ University Erlangen-N\"urnberg,\\ N\"agelsbachstr.\ 25, 91052 Erlangen, Germany 
}
\author{M.\ Halik}
\affiliation{
Organic Materials \& Devices, Institute of Polymer Materials, Department of Materials Science, \\
University Erlangen-N\"urnberg,\\
Martensstr.\ 7, D-91058 Erlangen, Germany
}
\author{T.\ Clark}
\affiliation{
Computer-Chemie-Centrum and Interdisciplinary Center for Molecular Materials,\\ University Erlangen-N\"urnberg,\\ N\"agelsbachstr.\ 25, 91052 Erlangen, Germany 
}
\author{M.\ Thoss}
\affiliation{
Institute for Theoretical Physics and Interdisciplinary Center for Molecular Materials, \\
University Erlangen-N\"urnberg,\\
Staudtstr.\ 7/B2, D-91058 Erlangen, Germany
}



\date{\today}

\begin{abstract}
We have investigated the conductance properties of $C_{60}$-containing 
self-assembled monolayers (SAMs), which are used in organic field-effect transistors, 
employing a combination of molecular-dynamics simulations, 
semiempirical electronic structure calculations and 
Landauer transport theory. The results reveal the close relation between
the transport characteristics and the structural and electronic properties of the SAM.
Furthermore, both local pathways of charge transport in the SAMs and the influence
of structural fluctuations are analyzed.

\end{abstract}


\maketitle

\section{Introduction}

Field-effect transistors (FETs) with thin films of organic $\pi$-conjugated materials as active semiconductor have developed towards a serious alternative for low-cost and flexible electronics on large areas.\cite{sekitani2,sekitani} The fact that charge transport in those organic films occurs essentially in the first molecular monolayer in close proximity to the dielectric interface \cite{tanase,mottaghi} makes a self-assembled monolayer field-effect transistor (SAMFET) a perfect device for studying charge-carrier transport properties in organic materials.\cite{brondijk} In SAMFETs, the active $\pi$-system is linked to a flexible insulating alkyl-chain that is covalently bound to a suitable surface. Thus, the $\pi$-systems form an almost 2D confined transport channel on molecular scale in thickness \cite{smits,schmaltz} and the n-alkyl linkers compose a hybrid gate dielectric together with the insulating anchor oxide ($AlO_x$).\cite{Klauk} Hole and electron transport have been demonstrated in SAMFETs making the concept a veritable approach in molecular scale electronics.\cite{smits,schmaltz,novak2,ringk}

%

The experimental advances have triggered theoretical studies of charge transport
in organic thin films within the context of SAMFETs.\cite{brondijk,jaeger} The theoretical description of
charge transport in these systems involves the principle problem
of accounting for structural flexibility and inhomogeneity. 
Thus, the characterization of the electronic structure must be combined with conformational
sampling, which in view of the complexity of the systems, represents a significant challenge. 
Furthermore, the mechanism of electron transport is {\em a priori} not
obvious. While dephasing due to both structural fluctuations and inelastic processes 
may favor hopping-type transport at higher temperatures, band-type coherent transport
can also contribute, in particular in systems with stronger intermolecular coupling and 
at lower temperatures.

In previous work, some of the authors investigated charge transport in
$C_{60}$-based SAMs using a combination of molecular-dynamics (MD) simulation, semiempirical electronic 
structure calculations and Monte Carlo (MC) transport studies.\cite{jaeger}
In agreement with experimental results,\cite{stubhan,novak} the study revealed the crucial influence of the 
morphology of the SAM on the transport properties.
In the present paper, we extend the previous study using a fully quantum mechanical description of transport.
Specifically, we employ Landauer transport theory,\cite{Landauer57} which provides a
coherent transport treatment based on elastic scattering mechanisms and is thus complementary 
to the MC-based investigations in Ref.\ \onlinecite{jaeger}.
Moreover, we use the theory of local currents \cite{pecchia} to investigate pathways of 
electron transport in the SAM and study the influence of structural fluctuations on
the transport properties.

The article is organized as follows: In Sec.\ 
\ref{sec:methods_setup}, we introduce  the SAMFET 
devices containing the layers of organic molecules,
 which are investigated throughout the paper. The 
theoretical methodology  is 
outlined in Sec.\ \ref{sec:methods_md} - \ref{sec:methods_loccurr},
including MD simulations, electronic
structure calculations, the Landauer transport approach and the theory of local currents.
In Sec.\ \ref{sec:results}, the results of the simulations obtained
for various SAMs, which differ in their structural properties, are presented. In particular, we
discuss the electronic structure and the transport properties
and their relation to the morphology of the SAMs.
In addition, local pathways for
charge transport in the SAMs are investigated and the influence of structural fluctuations on the 
transport properties of the SAMs is analyzed.

\begin{figure*}
	\centering
		\includegraphics[width=.95\textwidth]{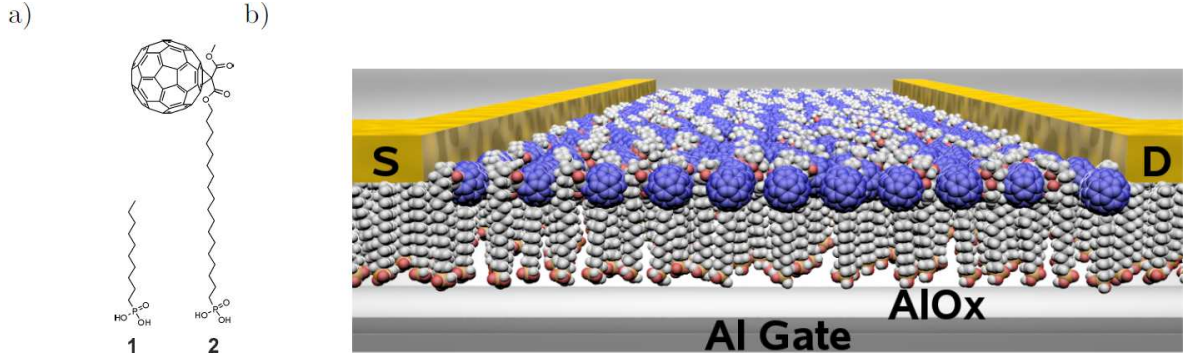}
		\caption{Schematic and chemical composition
 of SAMFET devices. (a) Chemical structure of the SAM 
forming phosphonic acids $C_{10}$-PA ({\bf 1}) 
and $C_{60}$-$C_{18}$-PA ({\bf 2}). 
(b) Schematic SAMFET setup with $Au$ source and 
drain electrodes, and $AlO_x$ gate dielectric between SAM and an $Al$ gate.}
\label{fig:device}
\end{figure*}


\section{Systems and Methods}  \label{sec:methods}

The theoretical methodology employed to simulate charge transport in 
SAMs involves three steps: (i) a characterization of the structure of
the SAMs with MD simulations, (ii) a determination of the electronic structure
using semiempirical molecular orbital (MO) theory, and (iii) transport calculations
within Landauer theory. In the following, the different methods are described. 
We begin with a brief introduction of the systems investigated and their use in SAMFET devices.

\subsection{Setup} \label{sec:methods_setup}

A scheme of the self-assembled monolayer field-effect
 transistor (SAMFET) device and its chemical 
composition is shown in Fig.\ \ref{fig:device}. The SAM is formed by fullerene-functionalized 
octadecyl-phosphonic acids (PAs) (in the following denoted by 
$C_{60}$-$C_{18}$-PA)
 and $C_{10}$-PA in various stoichiometric ratios. 
The SAM is separated from the aluminum gate electrode
 at the bottom by an $AlO_x$ layer, which together 
with the alkyl chains of the SAM builds the dielectric
 of the device.\cite{tbauer} Lithographically 
patterned gold (30 nm) is placed on top of the SAM, serving as
 source and drain electrodes.\cite{jaeger} The 
semiconducting $C_{60}$ head groups of the functionalized
 PA in the SAM form the active transistor channel in the device.
In this paper, we focus on the modeling of charge transport within 
the SAM. Therefore, the influence of a gate potential and 
the $AlO_x$ layer is not taken into account. The coupling to the
gold electrodes is described implicitely using self energies (see below).
 
\subsection{Molecular-Dynamics Simulations and Semiempirical MO Theory} \label{sec:methods_md}

The underlying conformational sampling of the self-assembled monolayer systems investigated in this publication is based on classical atomistic 
molecular-dynamics simulations published and described in detail 
previously.\cite{jaeger} Here, the SAM-forming molecules were initially aligned 
on a clean (0001) aluminum-oxide surface that was equilibrated prior to 
depositing phosphonic acids using an interatomic potential model 
parameterized by Sun et al..\cite{sunj} The parameters for the phosphonates are 
based on the general Amber force field (GAFF) \cite{wangj} and the MD simulations 
were performed with the program DL-POLY.\cite{todorovsmith} Details of the simulation 
are given in Ref.\ \onlinecite{jaeger}.

Following the MD simulations, an ensemble of snapshots was extracted and 
the $AlO_x$ substrate was removed for further processing. The snapshot 
geometries were then treated by semiempirical MO 
calculations using the restricted Hartree-Fock formalism and the AM1 
Hamiltonian.\cite{dewar} All semiempirical MO calculations were performed using 
the parallel EMPIRE program. \cite{empire}

\subsection{Model and Transport Theory} \label{sec:methods_transport}

To study charge transport through the SAM, we use an effective 
single-particle model with parameters determined by the
semiempirical MO calculations discussed above. Correspondingly, for a given 
snapshot of the structure of the SAM, the 
Hamiltonian of the SAM reads
\begin{equation}
H_S= \sum_j |\phi_j\rangle\epsilon_j \langle \phi_j|.
\end{equation}
Thereby, $\epsilon_j$ denotes the energy of an electron 
in the $j$th MO  $|\phi_j\rangle$.

Within the Landauer approach, transport through the SAM is described by
the transmission function  $t(E)$, which for an electron with energy
$E$ is given by 
\cite{Datta}
\begin{equation}
 t(E)={\rm tr}\left\lbrace\Gamma_RG_S\Gamma_LG_S^{\dagger}\right\rbrace. 
\label{eqn:1}
\end{equation}
Hereby, $G_S$ denotes the retarded Green's function of the SAM, 
\begin{equation}
 G_S(E)=\left[ E^+ - H_S-\Sigma_L(E)-\Sigma_L(E)\right]^{-1},
\end{equation}
which involves the self energies $\Sigma_{\alpha}(E)$
that describe the coupling of the SAM to the left and 
right electrodes ($\alpha=L,R$).
 The self energies are related to the width-function 
$\Gamma_{\alpha}(E)$, used in Eq.\ (\ref{eqn:1}), via 
\begin{equation}
 \Sigma_{\alpha}(E)= \Delta_{\alpha}(E)-\frac{i}{2}\Gamma_{\alpha}(E),
\end{equation}
where $\Delta_{\alpha}$ denotes the level-shift function.

The calculation of the self energy requires explicit modeling of the gold
electrodes and the gold-SAM interface.\cite{Benesch08,cuevasscheer} In this paper,  we
use a simpler strategy. We assume that the gold electrodes
couple preferentially to selected 
carbon atoms of the $C_{60}$ head groups at the
left and right boundaries of the SAM. This choice is suggested by the
geometrical structure of the SAM. Furthermore, we invoke the 
wide-band limit, where $\Gamma_{\alpha}$ can be approximated 
by a constant value and 
$\Delta_{\alpha}$ vanishes. The wide-band limit is a good 
approximation for gold electrodes.
Specifically, the matrix elements of the self energies $\Sigma_{\alpha}(E)$ 
in a local basis (represented by atomic orbitals $|\chi_{\nu}\rangle$) 
are given by
\begin{equation}
 (\Sigma_{\alpha}(E))_{\nu\nu}=-\frac{i}{2}(\Gamma_{\alpha})_{\nu\nu}
\end{equation}
 with $(\Gamma_{\alpha})_{\nu\nu}= 1 eV$
for orbitals $\nu$ corresponding to the outermost hexagon of Carbon atoms 
of the $C_{60}$ head groups at the left and right boundary of the SAM
and $(\Gamma_{\alpha})_{\nu\nu'}= 0$ otherwise.
The value $(\Gamma_{\alpha})_{\nu\nu}= 1$ eV respresents a reasonable choice 
for molecule-gold contacts. It is emphasized 
that the results
do not depend significantly on the exact value of this parameter, because 
the bottleneck for charge transport in the systems considered is the SAM
itself and not the SAM-gold contact.

Based on the transmission function $t(E)$, the electrical current through
the SAM is given by the expression
\begin{equation}
 I(V) = \frac{2e}{h}\int {\rm d}E \, t(E)(f_L(E)-f_R(E)),
\label{eqn:curr}
\end{equation}
where $f_{L/R}(E)$ denotes the Fermi function for the electrons in the
left/right lead. It is given by
\begin{equation}
f_{\alpha}(E)= \frac{1}{1+e^{(E-\mu_{\alpha})/k_BT}}
\end{equation}
with the Boltzmann constant $k_B$, the electrode temperature $T$ and their chemical potentials $\mu_{\alpha}$ ($\alpha=L,R$). For a symmetric voltage drop around the Fermi energy $E_F$, the chemical potentials are given by
\begin{equation}
 \mu_{\alpha}=E_F\pm \frac{eV}{2},
\end{equation}
where $V$ denotes the bias voltage. In the evaluation of the current, we assume that the transmission function
$t(E)$ depends only weakly on the bias voltage and use its value at 
zero bias voltage.

It is noted that Landauer transport theory, as used here,  
can only describe  coherent transport based on elastic 
scattering mechanisms. Dephasing and inelastic processes 
due to electron-phonon or
electron-electron interaction are neglected. The study of these 
processes requires an extension of the model
and more advances transport methods, which will be the topic of future work. 
The influence of structural fluctuation will be considered in Sec.\ 
\ref{sec:results_fluc}.

\subsection{Analysis of Transport Pathways Using Local Currents}  \label{sec:methods_loccurr}

Eqs.\ (\ref{eqn:1}) and (\ref{eqn:curr}) describe charge transport in terms
of the overall electrical current and transmission function of the complete SAM.
In order to have a more detailed description and, in particular, 
to analyze pathways
of charge transport in the SAM, we use  the method
 of local currents.\cite{Todorov,solomon}
As discussed in detail in Refs.\ \onlinecite{Todorov} and \onlinecite{solomon},
within this technique
the overall current $I$ through a surface perpendicular to the 
transport direction
is represented in terms of local currents via
\begin{equation}
 I=\sum_{m\in M_L}\sum_{n\in M_R} I_{mn},\label{eqn:Iloc}
\end{equation}
where $M_{L/R}$ denotes atomic sites left and right of the chosen 
surface, respectively.
As can be shown,\cite{solomon,pecchia} the local contributions 
to the current from site $n$ to $m$ are given by
\begin{equation}
 I_{mn}=\frac{2e}{h}\int \frac{dE}{2\pi}K_{mn}(E),
\label{eqn:localcurrs}
\end{equation}
where 

\begin{equation}
 K_{mn}(E)=\sum_{\nu\in m}\sum_{\mu\in n}
(V_{\nu\mu}G_{\mu\nu}^<(E)-V_{\mu\nu}G_{\nu\mu}^<(E)),
\label{eqn:2}
\end{equation}
and 
\begin{equation}
 G^{<}=(if_LG_S\Gamma_L G_S^{\dagger}
+if_RG_S\Gamma_R G_S^{\dagger}).
\end{equation}
Thereby, we have used a representation in the local atomic orbital basis
$|\chi_{\nu}\rangle$, where the Hamiltonian reads
\begin{equation}
 H_S= \sum_{\mu} |\chi_{\mu}\rangle\tilde\epsilon_{\mu} \langle \chi_{\mu}|+\sum_{\mu\ne\nu} |\chi_{\nu}\rangle V_{\nu\mu} \langle \chi_{\mu}|.
\end{equation}
Here, $\tilde\epsilon_{\mu}$ denotes the energy of an electron in orbital
$\mu$ and $V_{\nu\mu}$ the coupling constant between orbitals $\nu$ and $\mu$.
For temperature $T=0$, the expression
$\sum_{m\in M_L}\sum_{n\in M_R} K_{mn}$ can be 
identified with the transmission function, Eq.\ (\ref{eqn:1}), 
and thus $K_{mn}$ defines local contributions to the transmission
between pairs of atoms.

\begin{figure*}
	\centering
		\includegraphics[width=.95\textwidth]{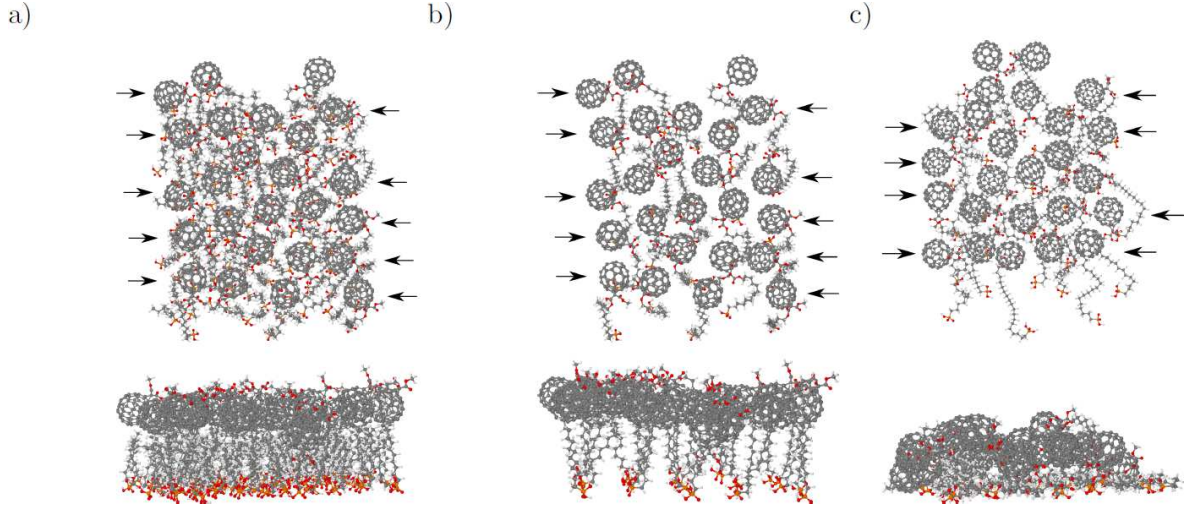}
		\caption{Top- and sideview of SAM consisting 
of 25 $C_{60}$-$C_{18}$-PA (a) mixed with 75 $C_{10}$-PA 
after a MD simulation time of 50 ns (b) without the $C_{10}$-PA,
 but with identical geometry as shown in (a),  
(c) Pure  25 $C_{60}$-$C_{18}$-PA SAM morphology as 
derived from a different simulation.}
\label{fig:systems}
\end{figure*}

In the applications considered below, we use a further coarse graining, where
the indices $m$ and $n$ in Eqs.\ (\ref{eqn:Iloc} - \ref{eqn:2}) refer not 
to single atoms but to the $C_{60}$-$C_{18}$- 
or $C_{10}$-PA moieties in the SAM. Consequently, the
sums in Eq.\ (\ref{eqn:2}) also extend  
over the orbitals of all atoms of each $C_{60}$-$C_{18}$- 
or $C_{10}$-PA moiety. This allows a more straightforward analysis 
of the pathways of charge transport through the SAM.

\section{Results and Discussion}\label{sec:results}

We have investigated three different SAM configurations 
as shown in Fig.\ \ref{fig:systems}, which differ both in the mixing ratio of
$C_{60}$-$C_{18}$ and $C_{10}$-PA groups and in their morphology.
The basic unit representing 
the SAM is a cell consisting of 25 $C_{60}$-$C_{18}$ PAs. 
Depending on the concentration of $C_{10}$-PA, the MD simulations 
reveal pronounced differences in the SAM morphologies. 
Fig. \ref{fig:systems} (a) and (c) show snapshots of a 25:75 
mixed SAM (containing, in addition, 75 $C_{10}$-PAs) and a pure SAM (without
$C_{10}$-PAs), respectively, after a MD simulation time of 50 ns. 
The pure SAM exhibits a considerably reduced thickness and an 
enhanced disorder of the $C_{60}$ head groups. Moreover, the $C_{60}$ 
head groups are much closer to the surface in the pure SAM and, 
therefore, favor leakage currents into the substrate. Monte Carlo (MC) 
simulations and experiments have proven this morphology 
to be less conductive and have suggested a mixing 
ratio in the range of 25:75 -- 50:50  to optimize the performance of 
the SAMFET devices. \cite{rumpel}

In addition, we have considered the SAM depicted in  
 Fig. \ref{fig:systems} (b), which was obtained from the
mixed SAM in panel (a) by removing the insulating $C_{10}$ alkyl chains
after the MD simulation. Although the geometrical structure of this
system is artificial, it can give insight into 
the role the $C_{10}$-PA plays in the transport through 
the layer. Note that the realistic morphology of the 
pure  SAM as derived from  MD simulations is the one shown in (c).

\subsection{Transport Properties and Molecular Orbital Analysis} \label{sec:results_mo}

\begin{figure}
 \includegraphics[width=1\linewidth]{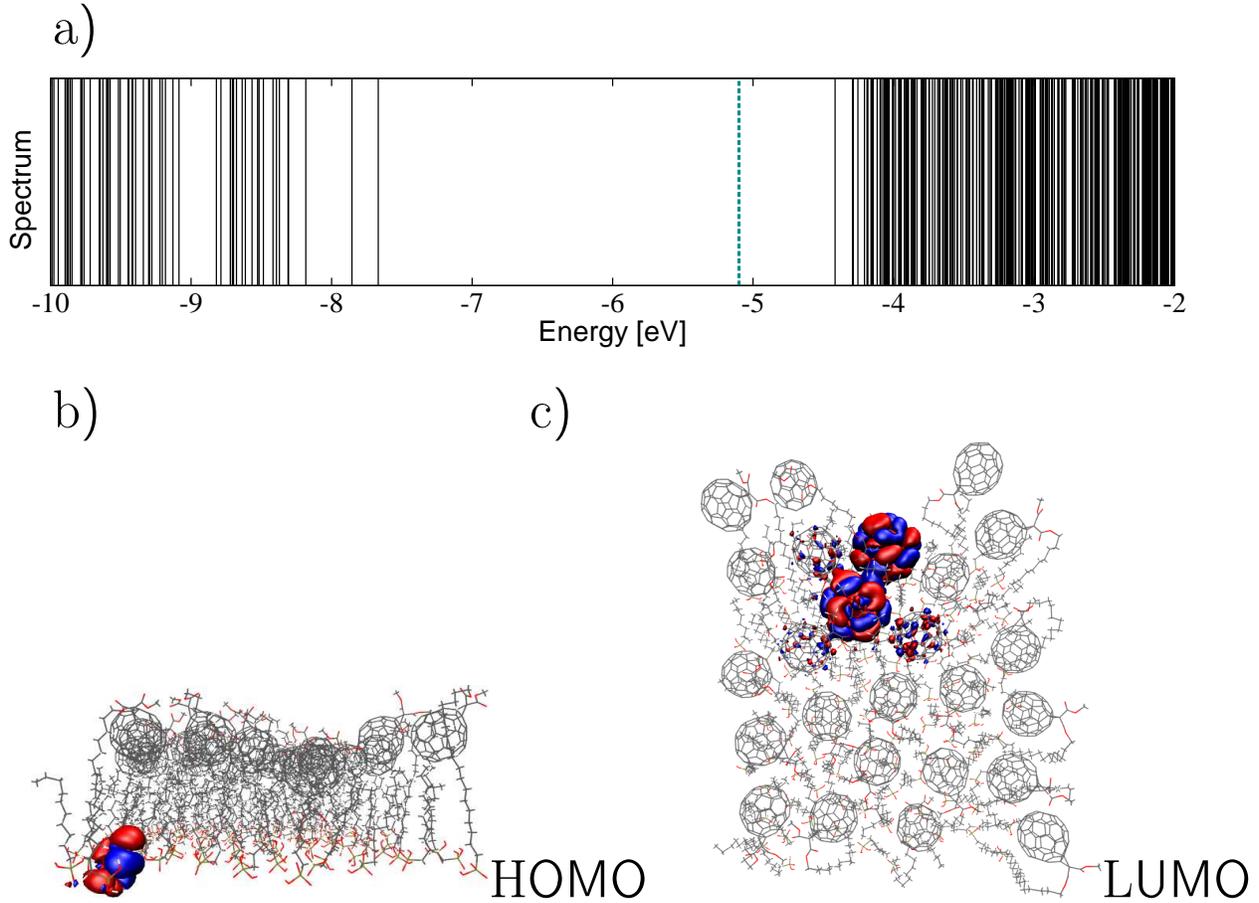}
\caption{ Electronic structure of the 25:75 mixed SAM. 
(a) Energy levels shown as black lines. The blue dashed line denotes the Fermi 
level of gold used in the transport calculations ($E_F=~-5.1$ eV). 
(b) highest occupied 
MO (c) lowest unoccupied MO. The MOs are plotted at isovalue $c=0.001$.}
\label{fig:spec}
\end{figure}

We first consider the 25:75 SAM depicted in Fig.\ \ref{fig:systems} (a). 
The energy spectrum of the MOs of this system in the range of [-10,-2] eV 
is shown in Fig \ref{fig:spec}.
\begin{figure*}
\includegraphics[width=.9\textwidth]{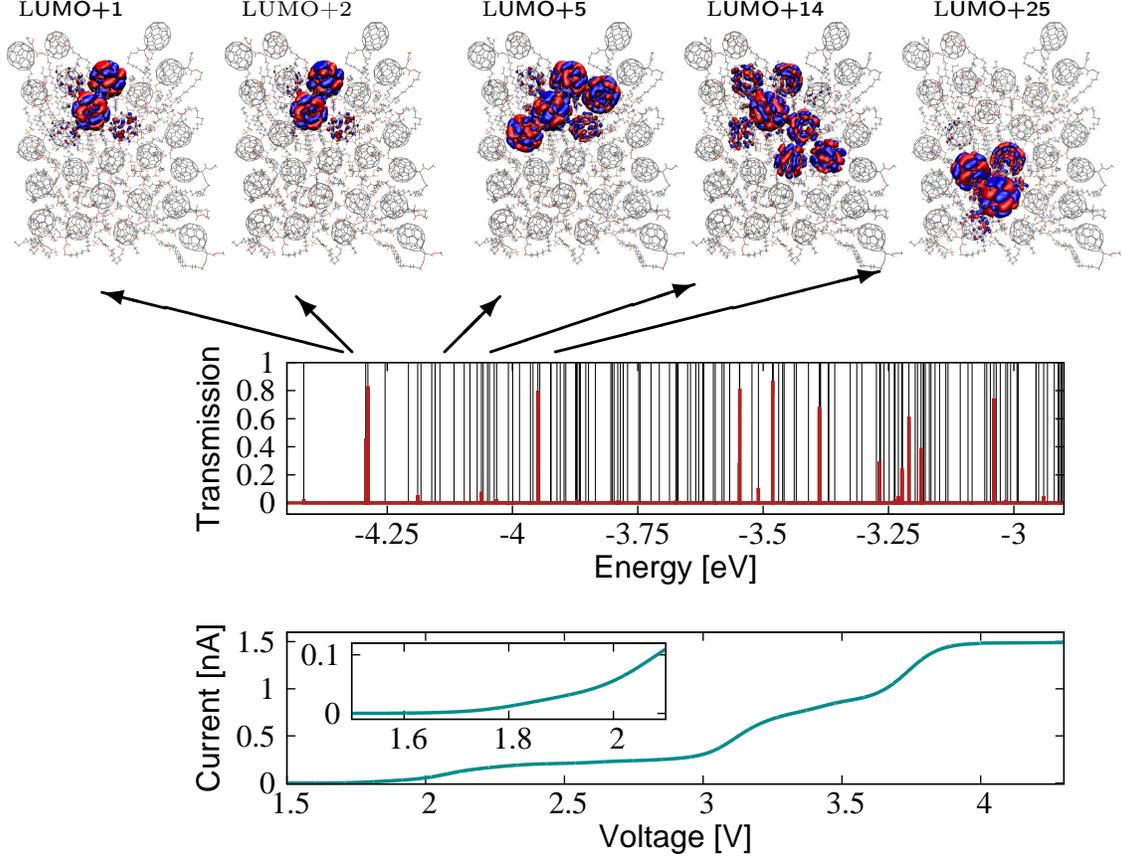}
\caption{Electronic structure and transport properties of the 25:75 mixed 
SAM Top: MO analysis, Center: Energy levels (black lines) 
and transmission function
(red lines), Bottom: Current-voltage characteristic}
\label{fig:moanalysis}
\end{figure*}
The spectrum exhibits a dense distribution of
energy levels and a HOMO-LUMO energy 
gap of 3.26 eV. In addition, the HOMO and the LUMO of the
SAM are depicted. Due to the pronounced structural disorder, 
all MOs are localized. The analysis 
reveals that the occupied orbitals close to the HOMO 
are mainly localized on the alkyl chains, while the unoccupied 
orbitals close to the LUMO are localized on the fullerenes. 
As fullerenes are electron conductors, only unoccupied energy states are 
relevant for transport processes. Therefore, in the 
simulations it is sufficient to take into account only 
the unoccupied part of the spectrum. 

 Fig.\ \ref{fig:moanalysis} shows the transmission function
 of the 25:75 SAM in an energy range close to the Fermi level 
(starting at $E_{\rm LUMO}$). In addition, the energy levels are 
depicted as black lines. The transmission
 function exhibits a series of narrow peaks, which can be associated to
 individual energy levels of the molecular layer. 
The small widths of the peaks are
 due to the strong localization of the 
corresponding molecular orbitals, which results in a weak coupling
to the electrodes. The 
current-voltage characteristic of the SAM at $T=300$ K, 
depicted in Fig. \ref{fig:moanalysis} (bottom), shows a 
non-Ohmic behavior. Thermally broadened steps in the current 
appear at voltages,
 where the chemical potential of the left electrode, 
$\mu_L=E_F+eV/2$, is in resonance with 
energies $E_i$ of the transmission peaks, i.e. at 
$V_i=2|E_F-E_i|$. 

As in the transmission function, the step structures
in the current can be associated to contributions from
individual MOs. For example, the rise of the current at $V\approx 1.65$ V
is due to contributions of LUMO+1 and LUMO+2. Although their 
transmission values are 
comparably large, their contributions to the current 
are small because of the small widths of their transmission
 peaks. The step at $V\approx 1.82$ V 
corresponds to LUMO+5, which 
exhibits a small transmission coefficient of $t=0.05$ 
but a much larger peak width.
At larger voltages, we 
find the contributions of several higher unoccupied 
MOs. The overall magnitude of the current is on the order of $nA$, 
which is in agreement with experimental data.\cite{jaeger} 

\subsection{Influence of Morphology and Mixing Ratio} \label{sec:results_mixed}

\begin{figure}
\includegraphics[width=.55\textwidth]{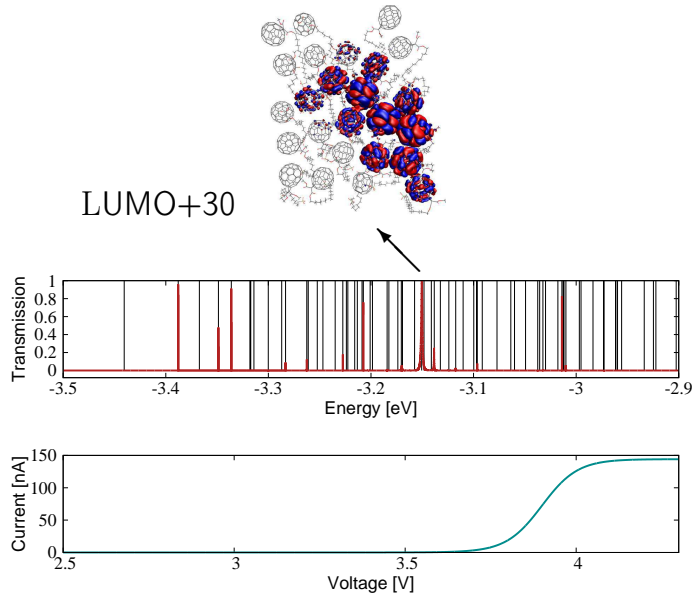}
\caption{Electronic structure and transport properties of the pure SAM with
  nonrelaxed geometry (see Fig.\ \ref{fig:systems} (b)) Top: 
MO analysis, Center: Energy levels (black lines) and transmission function (red lines), Bottom: Corresponding current}
\label{fig:moanalysis_wo}
\end{figure}

In order to explore the influence of the morphology of the SAMs 
and the mixing ratio of the SAM-forming molecules on the 
transport properties, we consider the 
pure $C_{60}$-$C_{18}$-PA SAMs with the two different 
geometries shown in Fig.\ \ref{fig:systems} (b) and (c).
We first focus on the pure SAM with the unrelaxed geometry,
which was obtained by removing the $C_{10}$-PAs groups from 
the 25:75 mixed SAM without further MD simulation.
The transmission function and the current-voltage characteristic
depicted in  Fig.\ \ref{fig:moanalysis_wo} 
reveal the pronounced influence of the 
$C_{10}$-PAs on the conductance properties of the SAMs.
Compared to the mixed SAM, we find that the LUMO 
of the pure systems is shifted by almost 1 eV to 
higher energies.  This and also a slight shift of the 
HOMO to lower values leads to an increase of the bandgap compared to
the mixed SAM. 
Due to the  larger bandgap, the onset of the current is shifted to
 higher voltages. However, the current magnitude is 
significantly larger than in the mixed SAM. 
The pronounced increase of the current at higher
voltages ($V \approx 3.8$ V) can be related to transport through
 LUMO+30. As illustrated in  Fig.\ \ref{fig:moanalysis_wo} (top panel),
LUMO+30 is a delocalized MO, which strongly 
couples to the electrodes resulting in a high current 
value of about 150 nA. Such delocalized MOs are
 not found in the mixed SAM. Thus, the presence of 
the $C_{10}$-PAs in the mixed SAM
appears to decrease the coupling 
between the fullerene head groups in the layer, which, in combination
with the disorder of the layer, results in the more localized orbitals
in the mixed SAM. Thus, in principle, the pure SAM with the geometry 
considered in Fig.\ \ref{fig:moanalysis_wo} has (at least for
higher voltages) better conductance properties than the mixed SAM.
It should be noted, though, that this is only the case for the
artificial geometry considered in Fig.\ \ref{fig:moanalysis_wo}.  
If the system is relaxed by further MD simulation, the morphology 
changes drastically to the geometry depicted in Fig.\ \ref{fig:systems} (c),
which will be considered next.

\begin{figure}
\includegraphics[width=.55\textwidth]{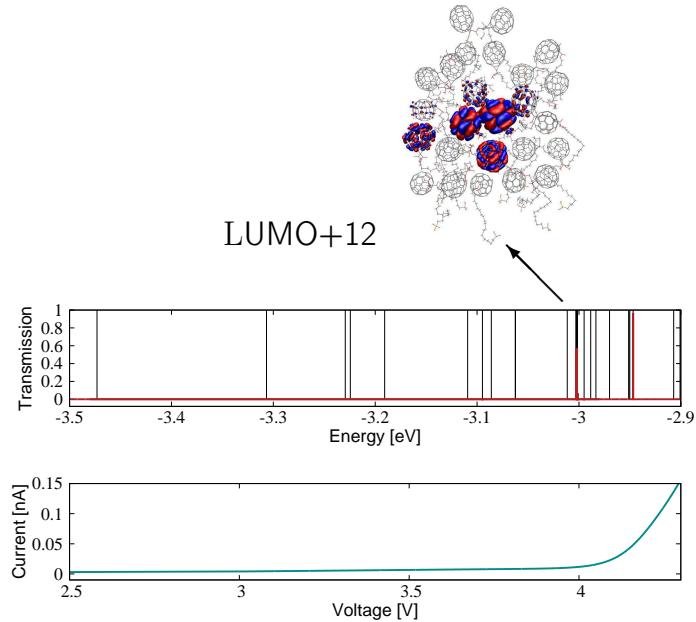}
\caption{Analysis of pure SAM (equilibrated) Top: 
MO analysis, Center: Energy levels (black lines) and transmission function (red lines), Bottom: Corresponding current}
\label{fig:moanalysis_flat}
\end{figure}

The electronic structure and conductance properties of
the pure SAM with the morphology depicted in 
Fig.\ \ref{fig:systems} (c) are shown in 
Fig.\ \ref{fig:moanalysis_flat}.
The results reveal that the almost flat and strongly disordered morphology
modifies the electronic structure profoundly and results
in a very small current, which confirms conclusions from
previous simulations and experiments.\cite{jaeger}
Evidently, due to the disorder, the 
$C_{60}$ head groups in the layer cannot form 
a conductive channel. The contributing MOs are 
strongly localized, leading to narrow peak structures
 in the transmission function.
Also, we find a shift of the LUMO to higher energies. 
The current onset is located at high voltages and 
has values about one order of magnitude smaller
than for the mixed SAM. These findings show 
that the poor conduction properties of pure 
$C_{60}$-$C_{18}$-PA SAMs can clearly be related
 to the decreased order in the molecular 
structure. Additionally, when used in 
SAMFETs, higher leakage currents can appear
 due to the smaller distance of the fullerenes 
to the substrate, as was indeed found in the transport measurements
\cite{burkhardt,jedaa}.

\subsection{Pathways for Electron Transport} \label{sec:results_paths}

So far, we have analyzed the conductance properties of the SAMs
in terms of the the overall transmission function and currents.
To obtain a more detailed understanding of the transport properties
and mechanisms,
we analyze the local pathways for electron transport through the SAM.
Previous investigations using Monte 
Carlo path searches based on a local electron affinity potential 
in the mixed SAM have suggested
 that there are specific pathways for efficient 
electron transport across the fullerenes. \cite{jaeger}
In the following we will extend this analysis based on the
quantum mechanical transport theory employed here using the
theory of local currents and transmissions outlined in Sec.\ 
\ref{sec:methods_loccurr}.

\begin{figure}[h!]
	\centering
\includegraphics[width=0.75\linewidth]{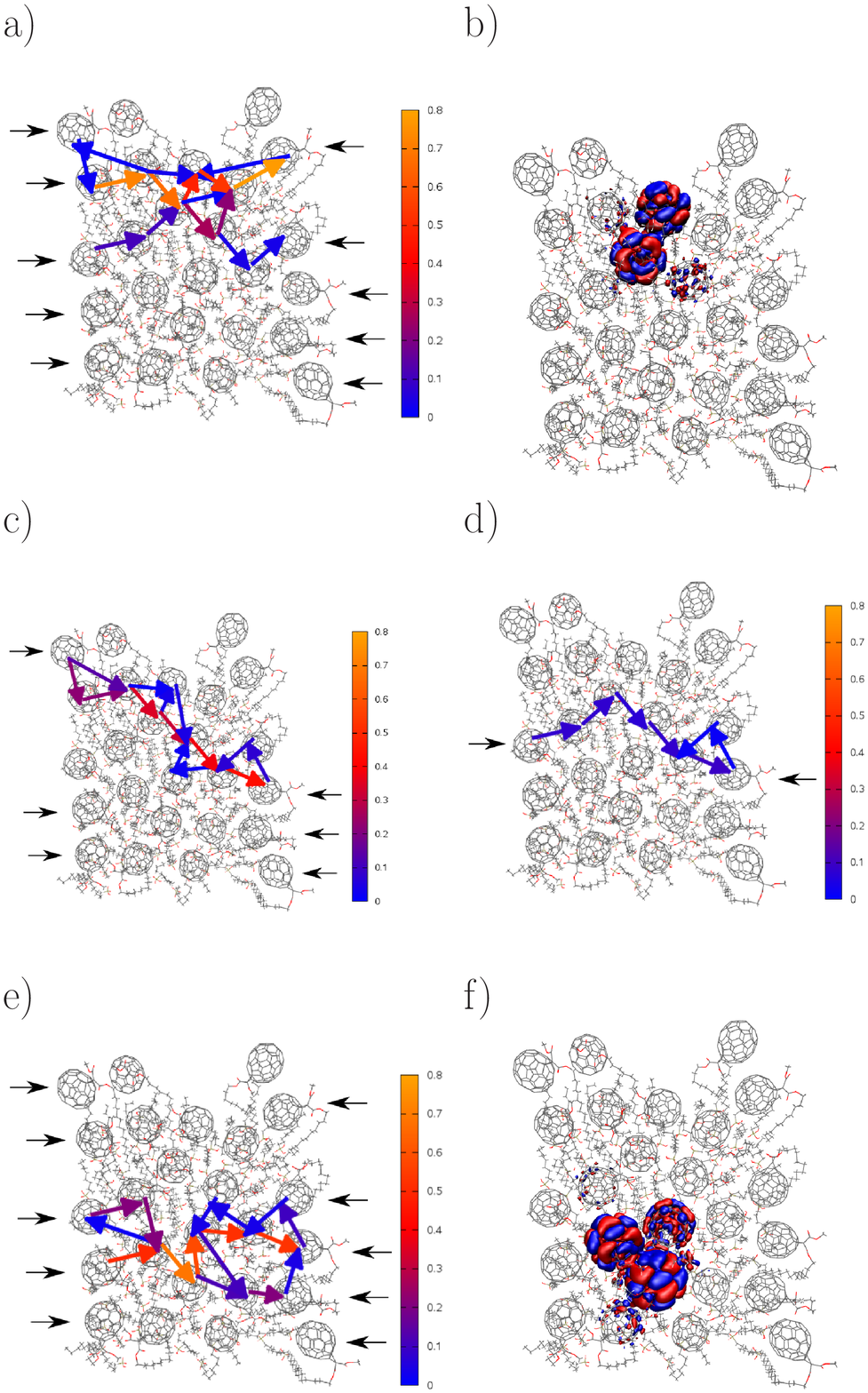}
\caption{Local transmissions between pairs of 
$C_{60}$-$C_{18}$-PAs and their relation to the 
MO structure. The colors of the arrows are ordered from blue 
to orange referring to an increasing transmission amplitude from 
0 to 0.8. Black arrows indicate the $C_{60}$ groups contacted by the
electrodes. (a,c,d) Local transmissions at 
$E_{\text{LUMO+2}}$ for different contact scenarios, 
(b) LUMO+2, (e) Local transmissions at $E_{\text{LUMO+25}}$, 
(f) LUMO+25} 
\label{fig:loctrans}
\end{figure}

We first consider the local transmission, defined
by Eq.\ (\ref{eqn:2}). Test calculations show that 
contributions from the insulating $C_{10}$-PA groups, e.g. 
from $C_{10}$-PA to $C_{60}$-$C_{18}$-PA  or
 between pairs of $C_{10}$-PAs, are very small. 
Therefore, we restrict the discussions to local 
contributions only between $C_{60}$-containing molecules.
The local transmissions were calculated at the energies $E$,
 where the total transmission (cf.\ Fig.\ \ref{fig:moanalysis}) exhibits its maxima. 
Each of these maxima can be related to a 
single MO in the spectrum. The results for energies $E_{\rm LUMO+2}$ (with
a total transmission of $t= 0.79$) and $E_{\rm LUMO+25}$ ($t= 0.69$) 
are shown in Fig. \ref{fig:loctrans}. In the plots, 
local transmissions between all $C_{60}$-$C_{18}$-PA 
molecules are depicted as arrows. The color code ranging from blue 
to orange indicates the local transmission value. 
An arrow is only drawn when this value exceeds 0.01. 
In this manner, a pathway for the electron through
the monolayer can be visualized. The shape of the pathway can clearly 
be related to the shape of the corresponding MO as 
shown, e.g.\ in Fig.\ \ref{fig:loctrans} (a,b) for 
LUMO+2. 
The non-vanishing local transmission components are 
mainly located on molecules which exhibit significant 
MO density. This is also illustrated for the transmission path 
at energy $E_{\rm LUMO+25}$ in Fig.\ \ref{fig:loctrans} (e,f).

\begin{figure*}
	\centering
\includegraphics[width=0.65\linewidth]{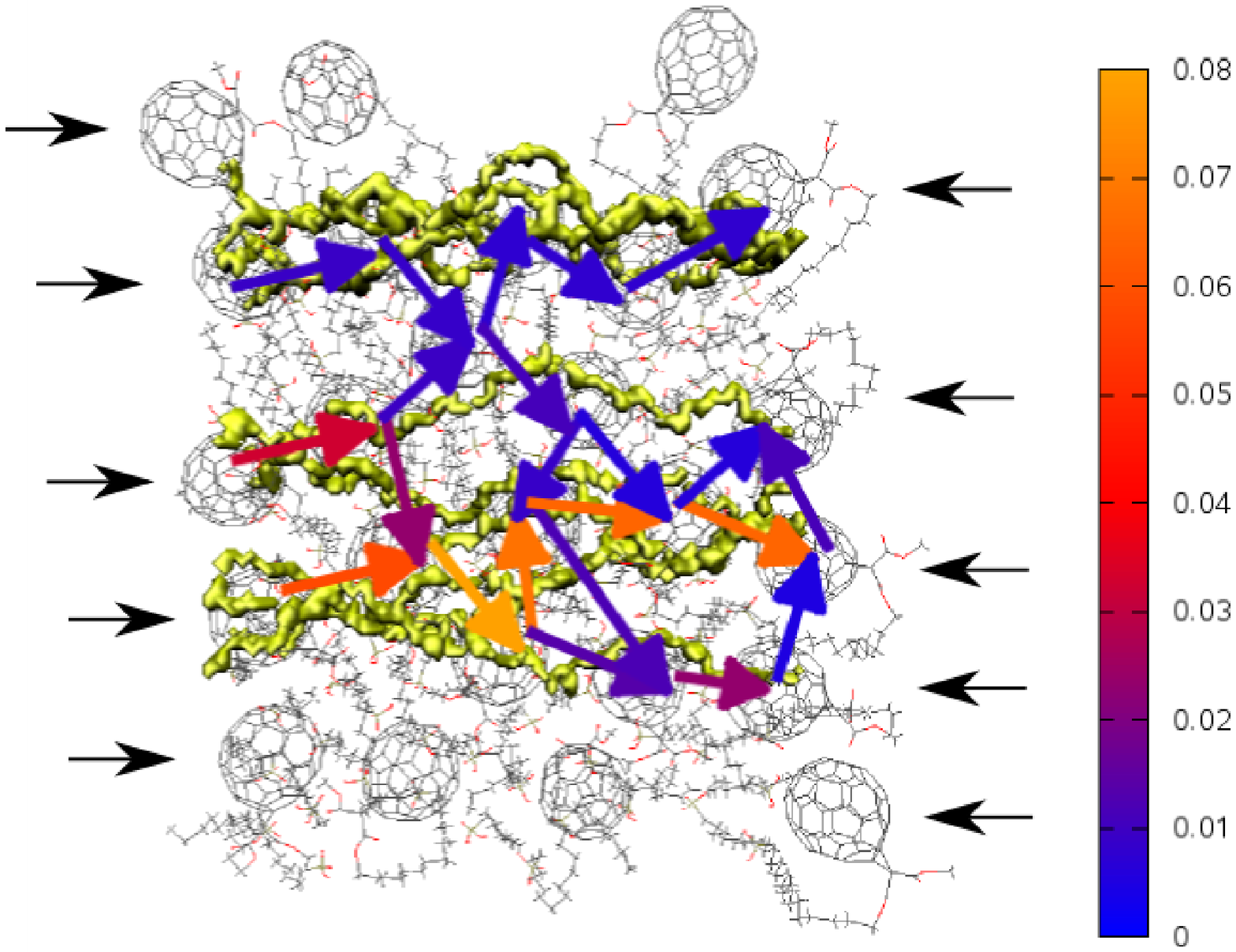}
\caption{Local currents at $V=2.4$ V (scale from blue to 
orange in nA) compared to MC transfer paths 
through the mixed SAM. The MC paths are shown as 
underlying yellow paths
and were calculated at $T= 2500$ K.}
\label{fig:mcpaths}
\end{figure*}

By integrating the local transmissions over energy, the 
local current is obtained, see Eq.\ (\ref{eqn:localcurrs}). 
This corresponds to a weighted sum of all local transmission paths 
contributing at a certain voltage. Tuning the bias voltage, 
the number of energy levels within the bias
window and thus the number of possible transport channels 
is increased. By plotting the local current components, we 
can visualize all channels which are open at a specific
 voltage. 
Fig.\ \ref{fig:mcpaths} shows the local currents at a bias 
voltage of $V=2.4$ V. The local currents are represented
as arrows between pairs of $C_{60}$-$C_{18}$ PAs, where 
the color of the arrows indicates the strength of the current. 
An arrow is only drawn if the current 
is larger than 0.01 nA. At the voltage considered, several transmission 
channels contribute. The dominating electron path corresponds to 
the contribution of LUMO+25 (cf.\ Fig.\ \ref{fig:loctrans} (e,f)). 
In the upper region of the 
SAM, smaller contributions of the lower unoccupied orbitals 
(especially LUMO+2,+5,+14, cf.\ Fig.\ \ref{fig:loctrans} (a)) are seen. 
There are also crossing components of the local currents
 between the different regions.
The pathways obtained based on the local current calculations
are compared in Fig.\ \ref{fig:mcpaths} to the results of
 MC paths searches (yellow colored density) performed using 
the local electron affinity energy map 
\cite{clark,jaeger} of the system. 
In the MC simulation, paths were initialized at the 
left boundary of the SAM such that they can start at all five 
outermost $C_{60}$ groups in agreement with the contact scenario used in the
local current calculation. In the MC simulation a temperature of
 $T= 2500$ K is used. Computationally this temperature can be attributed to the acceptance 
rate of the MC steps and physically it is related to a higher thermal 
energy of the electrons (285 meV) experimentally supplied by the 
operation voltage of the devices.
We find that the channels of high local currents agree well
with the MC results. There are some minor deviations.
For example, the MC simulation misses the currents that cross
between different regions. On the other hand, the MC simulation 
may also find paths localized in between the molecules, which
are by definition not present in the local current that
describes transitions between molecules.
The overall rather good agreement of the results obtained by two
quite different approaches supports the robustness of the results.

\subsection{Influence of Electrode-SAM Contact Geometry}

We have, furthermore, investigated the influence of 
the electrode-SAM contact geometry on the local 
transmission paths and the overall current.
This was achieved by varying the self energy (cf.\ Sec.\ \ref{sec:methods_transport}), 
which allows the  C$_{60}$ head groups to which the electrodes couple to be selected.
The influence of the contact scenario on transport 
is illustrated in Fig.\ \ref{fig:loctrans} 
for the local transmission at energy $E_{\rm LUMO+2}$.
The results show that the starting and end 
points of the pathways depend strongly on the 
C$_{60}$ head group the electrodes couple to. For example,
in Fig.\ \ref{fig:loctrans} (c) and (d), where a smaller number of C$_{60}$ where contacted, the
transmission values are significantly reduced.  Moreover 
we find that the transmission is strongly decreased 
if the electrodes couple only to molecules far away 
from those with MO density. 

\begin{figure}[h!]
	\centering
\includegraphics[width=0.95\linewidth]{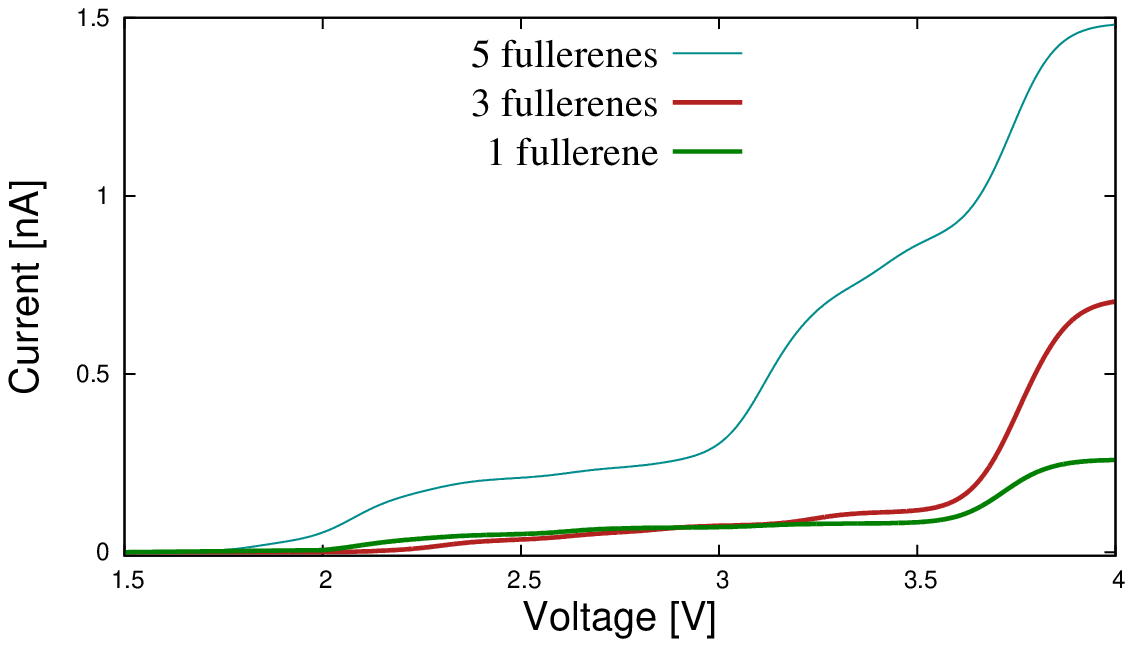}
\caption{ Comparison of currents for the different contact scenarios depicted in Fig. \ref{fig:loctrans} (a, c, d)}
\label{fig:currscont}
\end{figure}

The total current obtained for the three different contact 
scenarios depicted in 
Fig. \ref{fig:loctrans} (a, c, d) is shown in Fig.\ \ref{fig:currscont}. 
At larger voltages, where many transmission channels contribute, 
the overall current scales linearly with the 
number of molecules contacted. 
However, at smaller voltages this is not the case. 
For example, the current obtained for contact to only a single  
fullerene (cf.\ Fig.\ \ref{fig:loctrans} (d)) is 
larger than that for the scenario of Fig. \ref{fig:loctrans} 
(c) with three fullerenes in contact. 
This shows that for smaller voltage,
the number of contributing channels and thus the current 
magnitude depends sensitively on the specific electrode-molecule
contact scenario.

\subsection{Influence of Structural Fluctuations} \label{sec:results_fluc}

All results discussed so far were obtained
for a SAM with fixed geometry, i.e.\ a single snapshot taken at a fixed 
time of the MD simulation. However, structural fluctuations may 
play an important role for transport processes 
in the SAM. For instance, the formation of conductive 
pathways depends critically on the molecular arrangement, 
e.g.\ on the distance between the fullerene head groups. 

\begin{figure}
	\centering
\includegraphics[width=0.7\linewidth]{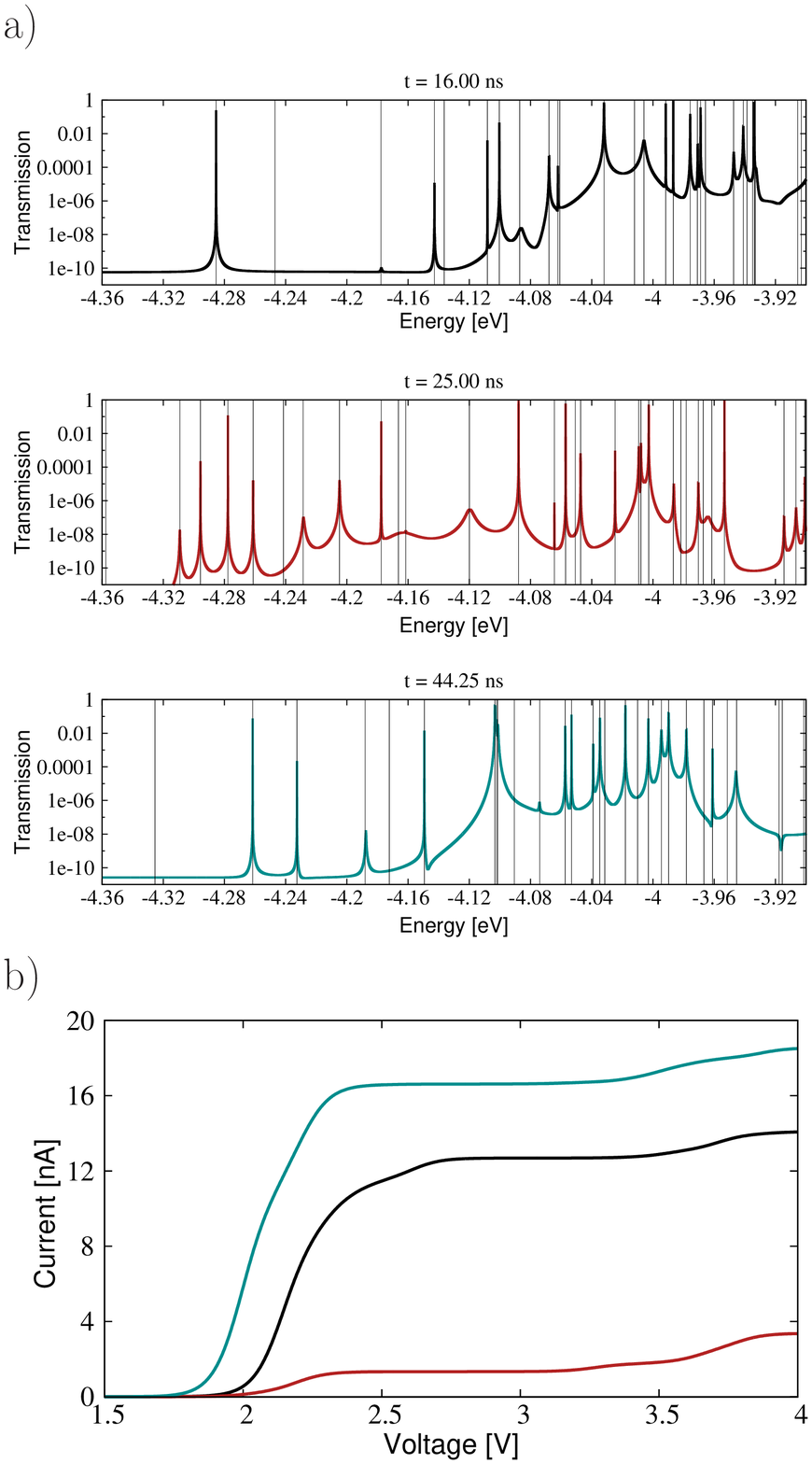}		
		\caption{Energy level spectrum, transmission function 
(a) and current-voltage characteristic (b) for different geometries of the 
SAM taken at three different 
times of the MD simulation.}
\label{fig:snaps}
\end{figure}

To study the importance of structural fluctuations, 
we have analyzed the electronic
structure and the  transport
properties for several snapshots taken
at different times along a MD trajectory.
Fig.\ \ref{fig:snaps} shows the energy spectrum, the transmission function and
the current-voltage characteristic for three representative examples,
 taken at $t=16$ ns, $t=25$ ns and $t=44.25$ ns.
The comparison of the  spectra for the different geometries 
reveals significant fluctuations of the energy levels. 
The energy of the LUMO, for example, fluctuates by about  
$0.1$ eV. As a result, the transmission functions change significantly, which,
in turn, results in fluctuations of the current by about one order of magnitude.
These examples show that structural fluctuations can influence the
electronic structure and the transport properties in the C$_{60}$-SAMs significantly.

A more detailed analysis of the influence of 
fluctuations on the transport
properties requires, in principle, a theoretical treatment beyond
the methodology employed here, using, e.g.\, time-dependent 
density matrix or nonequilibrium Green's function (NEGF) 
methods.\cite{Stefanucci2013} This will be the
the subject of future work.

\section{Conclusions}

We have investigated charge transport 
in $C_{60}$-based SAMs, which are used in SAMFET devises.
The theoretical study was based on a combination of MD simulations to characterize
the structure of the SAMs, semiempirical MO electronic structure calculation and 
Landauer transport theory.

The analysis shows that the transport characteristics can be related to
the structural and electronic properties of the SAMs. 
Due to the structural disorder of the SAMs, the
molecular orbitals are predominantly localized. As a result, the transmission characteristics
are dominated by narrow peaks, which can
be assigned to the specific molecular orbitals, and give rise to relatively low levels of electrical current,
in qualitative agreement with experiment.
A comparison of SAMs  with different mixing 
ratios of $C_{60}$  and alkene groups revealed, furthermore, the strong dependence of the 
transport properties on the layer composition. The 
conductivity depends crucially  on the formation 
of conductive transport channels across the $C_{60}$ 
head groups, which requires a certain degree of order. Extending previous work,\cite{novak,jaeger}
our studies show  that mixing  
$C_{60}$-$C_{18}$-PAs with $C_{10}$-PA increases the 
molecular order and facilitates transport. In SAMs consisting of pure
 $C_{60}$-$C_{18}$-PAs, on the other hand, the disordered molecular structure
 hinders the formation of conduction channels and results in a poor performance. 

We have also analyzed pathways of charge transport in the SAMs employing the theory of local currents.
The results demonstrate that the electrical current is dominated by few selected pathways,
which facilitate transport through the SAM. The specific contact scenario to the electrodes may influence the
selection of pathways, and thus affect the level of the electrical current.

Another important aspect of transport in organic SAMs is the 
effect of structural fluctuation on the transport properties.
Our results show that although the structural fluctuations within the SAM are relatively small, 
they have a significant effect on the transport properties and result in fluctuations of 
the current by about an order of magnitude. This is 
especially due to fluctuations in the alignment of the fullerenes.
To further examine the role of
fluctuations on the transport properties, 
a time-dependent treatment of charge transport 
\cite{Woiczikowski,croy} is required, using, e.g., 
time-dependent nonequilibrium Green's function (NEGF) theory.\cite{popescu} 
This may also open the door to a more complete theoretical treatment, including
inelastic processes due to electron-electron and electron-phonon interaction.

\section*{Acknowledgments}
This work was supported by the
the Deutsche Forschungsgemeinschaft (DFG) through the Cluster
of Excellence 'Engineering of Advanced Materials' (EAM) and
SFB 953. Generous allocation of
computing time at the computing centers in Erlangen (RRZE),
Munich (LRZ), and J\"ulich (JSC) is gratefully acknowledged.


%

\end{document}